\documentclass[english]{article}
\usepackage[T1]{fontenc}
\usepackage[latin9]{inputenc}
\usepackage{textcomp}
\usepackage{amsmath}
\usepackage{amssymb}
\usepackage{graphicx}
\usepackage{color}

\makeatletter
 \usepackage[a4paper,margin=1in]{geometry}
\usepackage{authblk}
\usepackage{cite}

\title{Site-specific recombinatorics: \textit{in situ} cellular barcoding with 
the Cre Lox system}
\author[1]{Tom S. Weber}
\author[2]{Mark Dukes}
\author[3]{Denise Miles}
\author[3]{Stefan Glaser}
\author[3]{Shalin Naik}
\author[1]{Ken R. Duffy}
\affil[1]{Hamilton Institute, Maynooth University, Ireland}
\affil[2]{University of Strathclyde, Glasgow, United Kingdom}
\affil[3]{The Walter and Eliza Hall Institute of Medical Research
\& The University of Melbourne, Parkville, Australia}

\usepackage{url}

\makeatother

\usepackage{babel}
\begin{document}
\title{Site-specific recombinatorics: \textit{in situ} cellular barcoding
with the Cre Lox system}
\maketitle
\begin{abstract}
Cellular barcoding is a significant, recently developed, biotechnology
tool that enables the familial identification of progeny of individual
cells in vivo.  Most existing approaches rely on ex vivo viral
transduction of cells with barcodes, followed by adoptive transfer
into an animal, which works well for some systems, but precludes
barcoding cells in their native environment, such as those inside
solid tissues.  With a view to overcoming this limitation, we propose
a new design for a genetic barcoding construct based on the Cre Lox
system that induces randomly created stable barcodes in cells in
situ by exploiting inherent sequence distance constraints during
site-specific recombination. Leveraging this previously unused
feature, we identify the cassette with maximal code diversity.  This
proves to be orders of magnitude higher than what is attainable
with previously considered Cre Lox barcoding approaches and is well
suited for its intended applications as it exceeds the number of
lymphocytes or hematopoietic progenitor cells in mice.  Moreover,
it can be built using established technology.
\end{abstract}

{\bf Keywords:} cell fate tracking; cellular barcoding; cre lox
system; DNA stochastic programme; combinatorial explosion.

\section*{Introduction}

The fate of the progeny of two seemingly identical cells can be
markedly distinct. Well studied examples include the immune system
and hematopoietic system, for which the extent of clonal expansion
and differentiation has been shown to vary greatly between cells
of the same phenotype
\cite{buchholz2013,Gerlach2010One,Verovskaya2013Heterogeneity,
Naik2013Diverse,perie15}.
Fate and expression heterogeneity at the single-cell level are also
apparent in other systems including the brain
\cite{Johnson2015Singlecell,Yagi2013Genetic,Zeisel2015Brain} and
cancers
\cite{NolanStevaux2013Measurement,Bhang2015Studying,Klauke2015Tracing}.
Whether this heterogeneity is due to the stochastic nature of
cellular decision making, reflects limitations in phenotyping, is
caused by external events, or a mixture of effects, is a subject
of active study
\cite{rohr2014single,Duffy2012Intracellular}.
As addressing this pivotal question through population-level analysis
is not possible, tools have been developed that facilitate monitoring
single cells and their offspring across generations.

Long-term fluorescence microscopy represents the most direct approach
to assess fate heterogeneity at the single-cell level. Studies
employing that technique are numerous
\cite{Hawkins09,Rieger2009,Gomes11,Giurumescu2012Quantitative,Duffy12,richards2
013},
and have revealed many significant features. Filming and tracking
of cell families \textit{in vitro} remains technically challenging,
is labor intensive, and only partially automatable
\cite{etzrodt2014quantitative,cohen2014extracting}. Despite significant
advances in the field, continuous tracking \textit{in vivo} is
confined to certain tissues, and time windows of up to twelve hours
for slowly migrating cells.

A radically different approach to long-term clonal monitoring is
to mark single cells with unique DNA tags via retroviral transduction,
a technique known as cellular barcoding
\cite{Gerlach2010One,Lu2011Tracking,Naik2014Cellular,Klauke2015Tracing}.
As tags are heritable, clonally related cells can be identified via
DNA sequencing. By tagging multi-potent cells of the hematopoietic
system and adoptively transferring them into irradiated mice, the
contribution of single stem cells to hematopoiesis has been
quantified
\cite{Lu2011Tracking,Naik2014Cellular}.
Amongst other discoveries, this has revealed 
heterogeneity in the collection of distinct cell types produced
from apparently equi-potent progenitors \cite{Naik2013Diverse,perie14,perie15}.
Current barcoding techniques are unsuitable for tagging cells
\textit{in vivo}, and typically require \textit{ex vivo} barcoding
followed by adoptive cell transfer \cite{Naik2014Cellular}.
This restricts its scope to cell types such as naive
lymphocytes and cancer cells, as well as hematopoietic stem and
progenitors which require perturbation of the new host, usually
irradiation, to enable them to engraft.

Ideally, a cellular barcoding system would inducibly mark cells
in their native environment, would be non-toxic,
permanent and heritable, barcodes would be easy to read with a
high-throughput technique, and the system would enable labeling
large numbers of cells with unique barcodes.  Two recently 
published studies address some of these points. Sun et al.
\cite{Sun2014Clonal} employed a Dox inducible form of
the Sleeping Beauty transposase to genetically tag stem cells
\textit{in situ}, and followed clonal dynamics during native
hematopoiesis in mice. There tags are the random
insertion site of an artificial transposon, which upon withdrawal
of Dox is relatively stable. A second \textit{in situ} cellular
barcoding system based on site-specific DNA recombination with the
Rci invertase has also been implemented 
\cite{Zador2012Sequencing,Wei2012Exactly,Peikon2014In}. Inspired
by the Brainbow mouse \cite{Livet2007Transgenic}, this system induces
a random barcode by stochastically shuffling a synthetic cassette
pre-integrated into the genome of a cell. The authors predicted
high code diversity from relatively small constructs (approx. $2$
kb) and demonstrated feasibility of random barcode generation in
Escherichia coli \cite{Peikon2014In}.

Each of those approaches elegantly overcome shortcomings of previous
systems by generating largely unique tags without
significant perturbation to the system of interest,
but some difficulties remain.  For barcode readout, the Sleeping
Beauty system requires whole-genome amplification technology and
three-arm-ligation-mediated PCR to efficiently amplify unknown
insertion sites. Furthermore, the random location of the transposon
may impact behavior of some barcoded clones and lead to biased data.
Moreover, some background transposon mobilization was detected,
subverting the stability of the barcodes. The Rci invertase based
system remains to be implemented outside bacteria. As with the
Sleeping Beauty transposase, the method requires tight temporal
control over Rci expression to make codes permanent.

Here we consider the Cre Lox system as a driver to induce
\textit{in situ} large numbers of distinct,
permanent, randomly determined barcodes from a series of tightly
spaced Lox sites. In contrast to the Brainbow construct
\cite{Livet2007Transgenic}, which relies on overlapping pairs of
incompatible Lox sites recombining randomly to one of several
stable DNA sequence configurations, our design exploits constraints
on the distance between Lox sites that arise during DNA loop
formation, a prerequisite for site-specific recombination
\cite{Hoess1985Formation,ringrose1999,Pinkney2012Capturing}. This
known feature has not previously been exploited, but is a crucial
design element for obtaining high barcode diversity. Employing
repeated usage of the same Lox site, code diversity is
solely restricted by cassette size and not, as in the Brainbow
construct, by the relatively small set of non-interacting Lox sites
\cite{Parrish2011BAC}. For a design without distance
constraints, the maximal diversity of stable barcodes creatable with the
Cre Lox system is of order $n$, where $n$ is the number
of Lox sites  \cite{Peikon2014In}, but with distance constraints
we establish that optimal barcode diversities of order $n^3$ are possible. 
Boosting this scaling with the four
incompatible Lox sites that have been reported in the literature
\cite{Parrish2011BAC} enables $10^{12}$ distinct codes of about 600 bp
each from a genetic construct as small as 2.5 kb.
In combination with the CreEr system \cite{Nagy2000Cre},
this is sufficient to inducibly barcode label
all naive CD8 T cells in a mouse \cite{Blattman2002Estimating}
or all nucleated cells in the bone marrow \cite{Colvin2004Murine}.
Desirable features are inherently part of the Lox barcode cassette
design, including: short and stable barcodes; a single barcode per
cell; and robust read-out.

\subsection*{Cre Lox biology }

Before introducing the Lox barcode cassette, we revisit Cre Lox
biology \cite{Sternberg1981Bacteriophage,Hamilton1984Sitespecific}.
Cre is a bacteriophage Pl recombinase that catalyzes site-specific
recombination between Lox sites. A Lox site is a 34 bp sequence
composed of two 13 bp palindromic flanking regions and an asymmetric
8 bp core region (Fig. \ref{fig:fig1} A). For recombination to occur,
four Cre proteins bind to the four palindromic regions of two Lox
sites and form a synaptic complex. A first pair of strand exchanges
leads to a Holliday junction intermediate \cite{Guo1997Structure}.
Isomerization of the intermediate then allows a second pair of
strand exchanges, and formation of the final recombinant product
\cite{Pinkney2012Capturing}. The DNA cleavage site is situated in
the asymmetric core region. If the Lox sites are on the same
chromosome, their interaction requires formation
of a DNA loop. If they have the same orientation (direct repeats),
recombination results in excision of the intervening sequence. If
Lox sites are in the opposite orientation to each other (inverted
repeats), the sequence between the sites is inverted, becoming its
reverse complement (Fig. \ref{fig:fig1} B).  Due to compatibility
with eukaryotes, the Cre Lox system has become an essential tool
in genetic engineering and a large array of transgenic mouse models
with inducible cell-type specific expression of Cre have been created
\cite{Nagy2000Cre}.

In \textit{in vitro} trials with Cre mediated Lox reactions, a sharp
decrease in recombination efficiency has been observed when the
sequence separating two Lox sites is less than 94 bp
\cite{Hoess1985Formation}. Recombination is still detectable at low
levels at 82 bp, but not at 80 bp where DNA stiffness appears to
prevent
DNA loop formation, and as a consequence
Lox site interaction. For the distinct, but similar, Flp/FR
system this minimal distance was established to be smaller \textit{in
vivo}, with interactions possible at 74bp \cite{ringrose1999}.
The existence of a minimal distance is one of the key features that
we exploit to make random barcodes stable, but in our proposed
design it will only prove necessary for it to be greater
than 44 bp.

\subsection*{Lox barcode cassettes}

In complete generality, a Lox barcode cassette is a series of Lox
sites interlaced with $n$ distinguishable DNA code elements of size
$m$ bp each. On Cre expression, code elements change orientation
and position, or are excised \cite{Wei2012Exactly}. Through Cre
mediated excision, the number of elements eventually decreases until
reaching a stable number (Fig. \ref{fig:fig1} C). Sequences that
have attained a stable number of code elements form size-stable
barcodes. A cassette's code diversity is the number of size-stable
barcodes that can be generated from the cassette via site-specific
recombination.

Our main result is a robust Lox cassette design that provably
maximizes code diversity. The design is robust to both sequencing
errors and to the minimal interaction distance between Lox sites.
The analysis that leads us to the design is provided in the Optimal
Design section. The identification of code element sequences that
avoid misclassification due to sequencing read errors then
follows. Finally, probabilistic aspects of code generation from an
optimal barcode cassette are explored via Monte Carlo simulation.
Lox cassettes with code elements of size 4 bp, higher order Lox
interactions, and the impact of transient Cre activation, are
considered in the discussion.

\subsubsection*{A robust cassette design that maximizes code diversity}

\begin{figure}
\centering\includegraphics[scale=0.7]{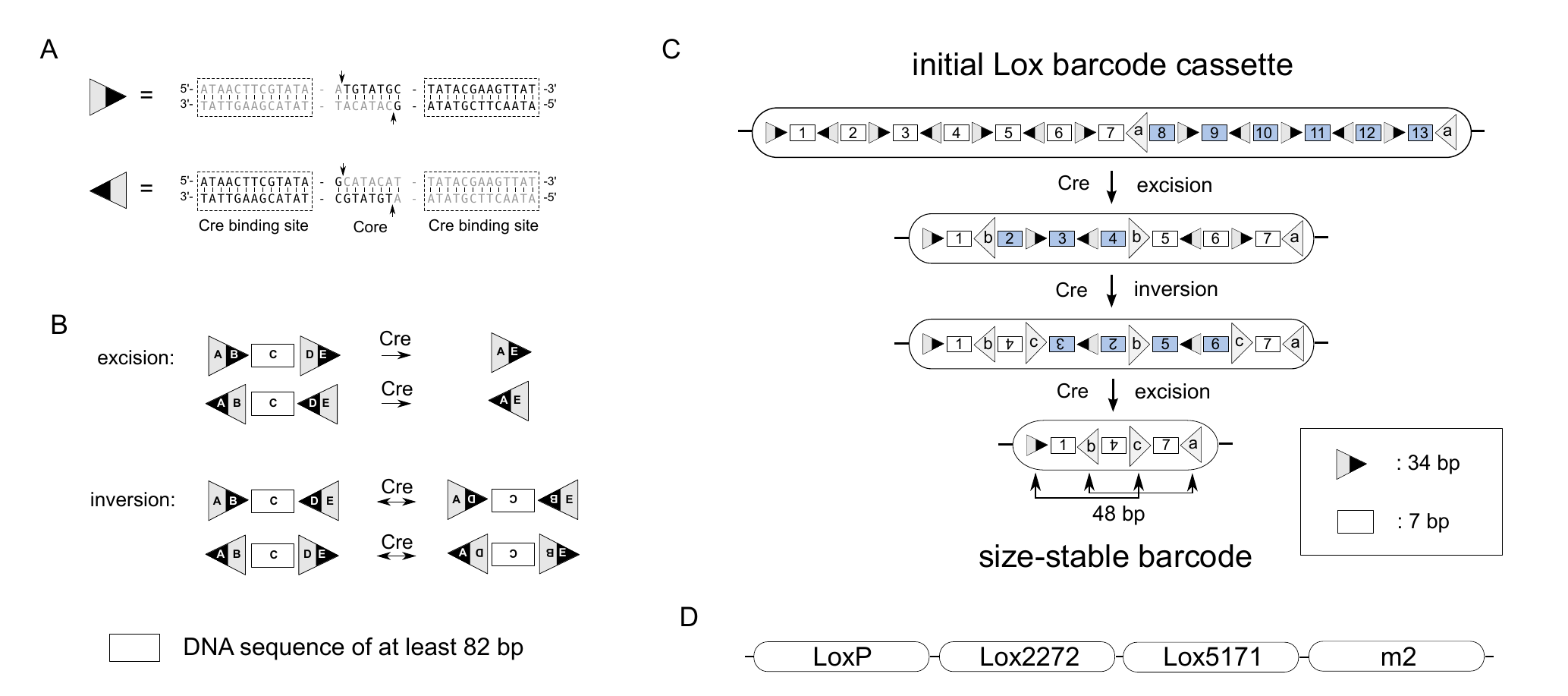}

\caption{Lox biology and Lox barcode cassette. \textbf{A}) Lox DNA
sequences. Lox sites are composed of two $13$ bp palindromic Cre
binding sites and an $8$ bp core (original LoxP sequence shown).
Cleavage sites in the core are indicated by arrows.  \textbf{B})
Cre mediated site-specific excision and inversion of a sequence
with a minimum of $82$ bp between two Lox sites on the same chromosome
\cite{Hoess1985Formation}.  If Lox sites are oriented in the same
direction, recombination excises the sequence, while if they are
oriented in opposite direction the sequence is inverted (i.e., the
reverse complement). \textbf{C}) An alternating Lox cassette with
13 elements of size 7 bp. To illustrate how barcodes are generated,
two excision and one inversion event are shown, creating a size-stable
barcode with three random elements.  Pairs of interacting Lox sites
are indicated by a, b, and c. Elements affected by recombination
have colored background. The barcode with three elements is size-stable
as Lox sites oriented in the same direction (arrows) are closer
than the minimal Lox interaction distance, precluding further
excision.  \textbf{D}) Four concatenated alternating Lox cassettes
of 13 elements each with poorly-interacting Lox site variants
\cite{Parrish2011BAC} result in a code diversity greater than
$10^{12}$. \label{fig:fig1}} \end{figure}

The optimal design will prove to have the orientation of both the
outmost, and any two consecutive, Lox sites inverted (Fig.
\ref{fig:fig1} C). Code elements between Lox sites are of size
longer than four bp, but shorter than 24 bp. The lower limit ensures
that elements can be chosen sufficiently distinctly to correct two
sequencing errors per element.  Due to the minimal Lox interaction
distance, the upper limit ensures that barcodes with three code
elements are size-stable.

The barcode diversity for this cassette design with $n$ code elements
under constitutive Cre expression will,
as established in the Optimal Design section,
transpire to be
\begin{align}
\frac{(n+1)(n-1)^{2}}{2}+(n+1)  =O(n^{3}),\label{eq:0}
\end{align}
which is maximal for code elements that are larger than four base
pairs.

A good compromise between cassette, robustness to sequencing errors and barcode
diversity is given by an alternating Lox cassette with 13 elements
of length 7 bp each as shown in Fig. \ref{fig:fig1}
C. The cassette is initially 567 bp long and generates a code
diversity of $1022$ barcodes.  After excisions and inversions,
size-stable barcodes are composed of either a single element or
three elements, with lengths 75 bp and 157 bp respectively, including
remaining non-interacting Lox sites. Concatenating four such cassettes
with poorly-interacting Lox variants (e.g. LoxP, Lox2272, Lox5171
and m2 \cite{Parrish2011BAC}, Fig. \ref{fig:fig1} D) yields a $2268$
bp construct with a size-stable code diversity of $1022^{4} \approx 10^{12}$. 

\subsubsection*{A practical implementation}
To implement Cre Lox barcoding in the mouse, one could cross mice
generated from embryonic stem cells that have been transduced with
the concatenated Lox barcoding cassettes described above onto a tamoxifen 
inducible cell-type specific CreEr expressing
background \cite{Nagy2000Cre}. An experiment is initiated by
administrating tamoxifen to the animal, which activates Cre and
induces generation of a barcode ($\leq 628$ bp) in each cell where Cre
becomes active.  Some time after activation, cells of interest are
harvested and sorted for specific phenotypes, and sequenced using
a next generation sequencing platform that produces read-lengths $>600$
bp. Cells originating from the same progenitor carry the same barcode
and this information can then used for scientific inference.  To
identify the frequent barcodes that are to be discarded in the
analysis (see the Barcode Distribution is Heterogeneous section),
in a control experiment large numbers of cells would be harvested
shortly after tamoxifen administration and sequenced.

\subsection*{Optimal design}

A simple upper bound on the barcode diversity of $k$ elements from
a cassette initially containing $n$ elements is the number of
possible outcomes when choosing $k$ from $n$ elements in arbitrary
order and orientation:
\begin{align*}
{n \choose k} k! 2^k = \frac{2^{k}n!}{(n-k)!}.
\end{align*}
Although loose, it will become clear that it captures the dominant
growth, $O(n^{k})$, indicating the importance of $k$ in generating
barcode diversity and motivating a closer look at how cassette designs
influence it.

For what follows, we introduce some terminology: a cassette is
alternating if the orientation of any two consecutive Lox sites
is inverted (Fig. \ref{fig:fig1} C); outermost Lox sites are termed
flanking Lox sites; and flanking sites are direct or inverted
if they have the same or opposite orientation, respectively.

\subsubsection*{Code diversity is determined by code element length and orienta
tion
of flanking sites}

Cre recombination requires a minimal distance between the interacting
Lox sites. In what follows we assume that the minimal distance for
Lox interaction is 82 bp, but our results will be robust for any
minimal interaction distance greater than 
44 bp.

\begin{figure}
\centering\includegraphics[scale=0.7]{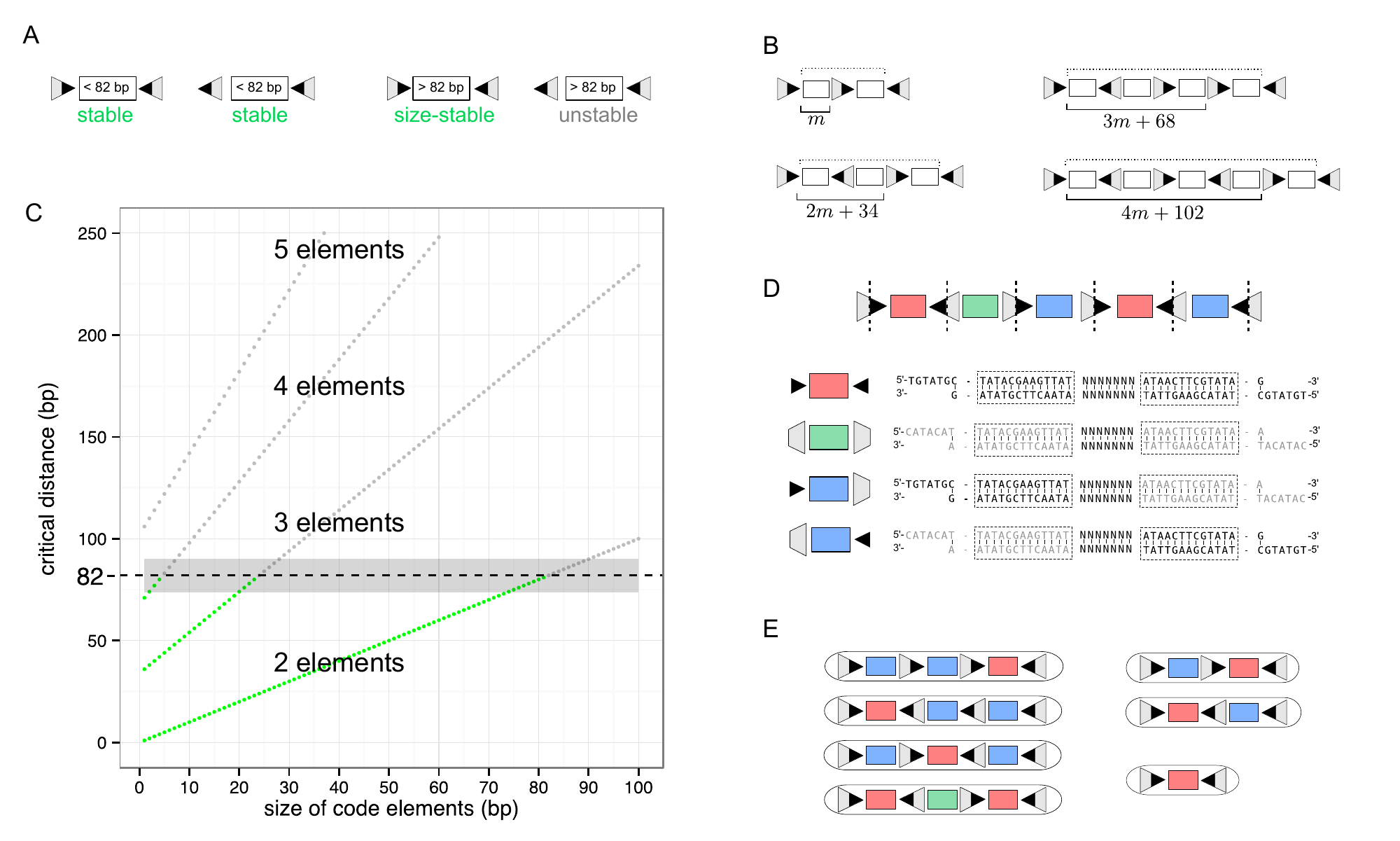}

\caption{Barcode stability and code diversity. \textbf{A}) Size-stability 
of barcodes with a single element depends on the length of
the sequence between the flanking sites and their relative orientation.
\textbf{B}) Critical distances of barcodes of different sizes from
a cassette with inverted flanking sites. Dotted lines show the critical
distance if flanking sites are oriented in the same direction. \textbf{C})
Stability of barcodes from $2$ to $5$ elements for a Lox barcode
cassette with inverted flanking sites. If the critical distance surpasses
the minimal distance, stable codes (green) become unstable
(gray). Barcodes of size three and four are unstable if $m\geq24$
and $m\geq5$ respectively, while codes of size five are always unstable.
The gray interval illustrates potential uncertainty in the estimate
of the minimal interaction distance. \textbf{D}) Sequences between
Lox cleavage sites represent the fundamental building blocks of
the barcode cassette. There are two with inverted Lox repeats
(red, green) and two direct Lox repeats (blue) types of blocks.
In the example, code elements are of size 7 bp and N denotes an
arbitrary base. \textbf{E}) For a cassette with inverted flanking
sites pointing at each other and $5\leq m<24$, four block compositions
are possible ($\{k_{r},k_{g},k_{b}\}$): two for barcodes of size
three (three $\{1,0,2\}$ and one $\{2,1,0\}$), one for barcodes
of size two (two $\{1,0,1\}$) and one for barcodes of size one
(one $\{1,0,0\}$).\label{fig:fig2}}

\end{figure}

To understand how a minimal Lox-Lox interaction distance and
cassette design determine size-stable barcodes and code diversity,
we start with the simplest case, a barcode with a single code element
(Fig. \ref{fig:fig2} A). If the code element is less than 82 bp,
the barcode is size-stable irrespective of the orientation of its
flanking sites. If the element is larger than 82 bp, the code is
only size-stable if the flanking sites are inverted as excision
will remove the element.

For a barcode with two elements, the sequence between the flanking
sites contains an additional element and a Lox site (34 bp), giving
a sequence of $2m+34$ bp. If the flanking sites have the same
orientation, the barcode is size-stable if $2m+34<82$ bp, hence if $m<24$
bp. If they are in opposite orientation, excisions can only occur 
if flanking sites interact with the middle Lox site, and $m<82$
bp is sufficient for stability (Fig. \ref{fig:fig2} B). For given
$m$, in general if there exists a barcode of size $k$ with direct
flanking sites, a barcode with $k+1$ elements is possible that has
inverted flanking sites. Thus $m$ and the orientation of the flanking
sites are critical features that determine the maximum $k$.

In Fig. \ref{fig:fig2} C, the stability of barcodes with $k\in\{2,3,4,5\}$
is shown as a function of $m$ for a cassette with inverted flanking
sites. The stability depends on a critical distance, i.e., the
largest distance between two Lox sites in the barcode that is, or
can be brought into, the same orientation via recombination. As
shown, barcodes of size three and four become unstable if $m\geq24$
bp and $m\geq5$ bp, respectively, while barcodes of size five or
greater are always unstable.

Orientation of a cassette's flanking sites is immutable under
recombination. Therefore cassettes with direct and inverted flanking
sites generate barcodes with direct and inverted flanking sites
only. Having seen that maximal code diversity grows as $O(n^{k})$,
and that having inverted flanking sites relative to direct ones
increases the maximum size of barcodes by one, it follows that the
diversity for cassettes with inverted flanking sites is of the order
$O(n^{k+1})$. Inverted flanking sites are thus superior in terms
of code diversity and are an essential design decision.

Optimality regarding the size of the elements, $m$, is more intricate.
For $m<5$, the maximum size of barcodes is four elements, and
according to the formula above, their diversity grows as $O(n^{4})$.
The stability of barcodes with four elements is, however, sensitive
to the minimal distance estimate (the gray interval in Fig.
\ref{fig:fig2} C). In addition, the short length of code elements
limits error correction, a point revisited later. Thus we focus on
cassettes in the regime $5\textrm{ bp}\leq m<24$ bp, which generate
error-robust barcodes of up to size three and a code diversity that
is insensitive to the reported minimal Lox interaction distance.

\subsubsection*{Alternating Lox cassettes with inverted flanking sites maximize
code diversity}

For the orientation of the remaining Lox sites we prove, via a two-step
strategy, that the alternating design produces maximal
code diversity. First we derive a refined upper bound for the
diversity that takes into account the structure of the Lox cassette,
but ignores constraints imposed by the recombination process. We
then show that alternating Lox cassettes with inverted flanking
sites and $n\geq7$ elements are unconstrained in terms of barcode
generation via sequential recombination events, thus achieving this
upper bound.

\subsubsection*{An upper bound for Lox barcode diversity}

During Cre induced recombination, Cre proteins cleave the core
region of the interacting Lox sites asymmetrically
\cite{Pinkney2012Capturing}. The sequences between subsequent
cleavage sites are not affected by Cre and represent the fundamental
building blocks of the Lox barcode cassette. Each block contains a code
element and half a Lox site on each side.

Depending on the orientation of the Lox sites, there are four
possible types of blocks (Fig. \ref{fig:fig2} D). Three colours
have been used to code these: red, green and blue. By definition,
the reverse complement of a block is of the same colour class. In
contrast to blue blocks, red and green blocks have their Lox cores
cleaved in a way such that their flanking Lox sites are unchanged
after inversion, while the intervening sequence is reverse-complemented.

Blocks are similar to the concept of units in
\cite{Wei2012Exactly}, introduced to derive expressions for the
total number of sequences, stable or unstable, generated from a Lox
cassette where all $(n+1)$ sites can interact. Their analysis implies
$m>82$ and a code diversity of order $n$. Quite distinctly, here
we focus on enumerating size-stable sequences that arise in the
regime $5\textrm{ bp}\leq m<24$ bp with code diversities of order
$n^3$.

Stable codes are necessarily made of blocks from the initial cassette,
and as shown in Fig. \ref{fig:fig2} E, their composition in terms
of block colors is prescribed. Letting $n_{r}$, $n_{g}$, and $n_{b}$
be the number of red, green, and blue blocks in the initial cassette
with $n$ elements, an upper bound on the number of possible barcodes
of size $k$ with $k_r$ red, $k_g$ green and $k_b$ blue blocks is
the number of possible outcomes when choosing
$k_{r}$, $k_{g}$ and $k_{b}$ from $n_{r}$, $n_{g}$ and $n_{b}$
elements in arbitrary order:
\begin{align*}
k_{r}!\dbinom{n_{r}}{k_{r}}k_{g}!\dbinom{n_{g}}{k_{g}}k_{b}!\dbinom{n_{b}}{k_{b
}}2^{k_{r}+k_{g}},
\end{align*}
where $n_{r}+n_{g}+n_{b}=n$ and $k_{r}+k_{g}+k_{b}=k$. 
The additional factor $2^{k_r+k_g}$ arises as there
are two valid orientations of every code element of a red and green
block after recombination. Conditioned on $n_{r}$, $n_{g}$, and
$n_{b}$, to derive an upper bound for a cassette's diversity, we
add the numbers for the four possible stable
barcode configurations of $k_{r}$, $k_{g}$, and $k_{b}$
(Fig. \ref{fig:fig2} E), taking into account
that certain configurations appear more than once (e.g.
the configurations with one red and two blue blocks appears three
times). Using the expression above for
each of the four configurations,
for $5$ bp $\leq m<24$ bp, and cassettes with inverted
flanking sites pointing at each other (the opposite case is similar)
this yields, 
\begin{align*} 
	3\left(1!{n_r \choose 1}2!{n_b\choose 2}\right)2^{1}
	+1\left(2!\dbinom{n_{r}}{2}1!\dbinom{n_{g}}{1}\right)2^{2+1}
	+2\left(1!\dbinom{n_{r}}{1}1!\dbinom{n_{b}}{1}\right)2^{1}
	+1\left(1!\dbinom{n_{r}}{1}\right)2^{1}.
\end{align*} 
By construction, $n_{g}=n_{r}-1$, and since
$n_{b}=n-2n_{r}+1$, substituting the respective
terms leads to an expression that is a function of $n$ and $n_r$
alone. For given $n$ odd, this reduces the task of finding the
optimal cassette design to an explicitly solvable one-dimensional
optimization problem: 
\begin{align*}
\underset{n_r}{\arg\max}\quad 32 n_r^3 - 12 (2 n + 3 ) n_r^2 + (6
n^2 + 10 n + 14) n_r
  \quad\textrm{ for }\quad n_r \leq \frac{n+1}{2}.  
\end{align*}
For $n\geq5$, the global maximum is achieved at
the boundary $n_{r}=(n+1)/2$. This implies $n_{b}=0$, and a global
upper diversity bound of $(n+1)(n-1)^{2}+(n+1)$, of order $O(n^{3})$.
It is easily verified that $n_{b}=0$ is only possible if the cassette
design is alternating and n is odd, which implies the flanking sites
are inverted.

\subsubsection*{Alternating Lox cassette design achieves the upper diversity bo
und}

For an alternating cassette design, achieving the code diversity
upper bound requires complete freedom in code generation via
recombination events. By construction, we show that this is the
case if $n\geq7$.

Consider an alternating cassette with five elements and $m\geq5$
bp, and recombination events that do not alter the size of the
cassette (i.e., inversions). First note that red blocks in position
three and five can move into the first position via a single
recombination event. Furthermore, a red block in position one can
be inverted by first moving to position three, then to five, and
back again. A straight-forward recipe to create an arbitrary code
made of a single red block is then to: i) move the block into the
first position (if required); ii) change its orientation (if
required); and finally iii) excise the remaining blocks.

Similarly, to generate an arbitrary code composed of a red and a
green block from an alternating cassette with six elements, we can
perform steps i) and ii). Then we apply the same procedure to the
green blocks, leaving the first block untouched. This results in
the first two blocks of the cassette being the desired
code. To generate the size-stable code, elements that are not part
of the code are excised.

Finally, for a cassette with seven elements, sequentially following
the recipe given above, the first three blocks can be populated such
that they match any possible code before excising the remaining blocks.
This shows that any possible code of size one to three can be created
via Lox recombination if the cassette is alternating, $n\geq7$,
$m\geq5$ bp, and flanking sites are inverted. 

Under constitutive Cre expression, barcodes with three elements can
still undergo inversions via the flanking sites, which reduces their
code diversity by a factor of two. The code diversity is therefore
that given in Eq. \eqref{eq:0}.

\section*{Design of code element sequences}

That barcodes generated from a Lox cassette are pre-defined in terms
of sequence and position in the genome represents an advantage over
barcoding systems that rely on insertion site analysis for barcode
readout \cite{Sun2014Clonal,Bystrykh2012Counting}.  If codes-reading
was error-free, choosing code elements of a particular color (Fig.
\ref{fig:fig2} D) from a set of sequences that differ at least by
one bp pair in both orientations would be sufficient. The maximum
number of such elements is $(4^{m}-4^{m/2})/2$ and $4^{m}/2$ for
$m$ even or odd, respectively, which is large even for small $m$.

In order to be perfectly robust to $j$ read errors via nearest-neighbor
match, all pairs of elements of a given color need to differ by a
Hamming distance of at least $2j+1$ bp \cite{Cover1991Elements}.
The size of the sets of elements that meet this condition quickly
decreases with increasing $j$ (see Fig.  \ref{fig:fig3} A for
numerical estimates). To ensure correction of two sequencing errors
requires $m\geq5$ bp.

Assuming that sequencing errors arise independently and error rates
are identical for all bases, the number of read sequencing errors
in a code element of size $m$ is Binomial with the error probability
per bp \cite{Li2008Mapping}. Any element that has $j$ or less errors
will be classified correctly by nearest-neighbor matching. The
probability of more than $j$ errors gives an upper bound for the
expected proportion of misclassified code elements. Fig. \ref{fig:fig3}
B shows this for elements of size $m=7$ bp as a function of the
minimal distance and the read error rates for next-generation
sequencing platforms \cite{Ross2013Characterizing}. Different symbols
indicate different sequence data. Even for low-fidelity platforms
like Pacific Bioscience single molecule real time sequencing, a
minimal distance of five bp results in less than ten misclassified
elements per million.

\section*{Probabilistic features of optimal Lox cassettes}

In this section we explore stochastic features of the optimal design,
specifically the probabilities to generate each of the final codes
and the number of recombination events that are needed to create
size-stable codes. For the analysis, we make two assumptions: first,
all interactions with Lox sites that are at least 82 bp apart are
equally likely; second, recombination events occur sequentially and
independently.

\subsubsection*{Barcode distribution is heterogeneous}

Size-stable barcodes of a Lox cassette are randomly generated and
not all codes are equally likely. Although an analytical expression
for the probability mass function of final codes is not available,
stochastic simulations enable us to study properties of practical
importance such as the probability of generating a code more than
once. Ensuring this probability is low is important in practice
because progeny of two cells that independently generate the same
code will be confounded as pertaining to the same clone.

\begin{figure}
\centering \includegraphics[scale=0.65]{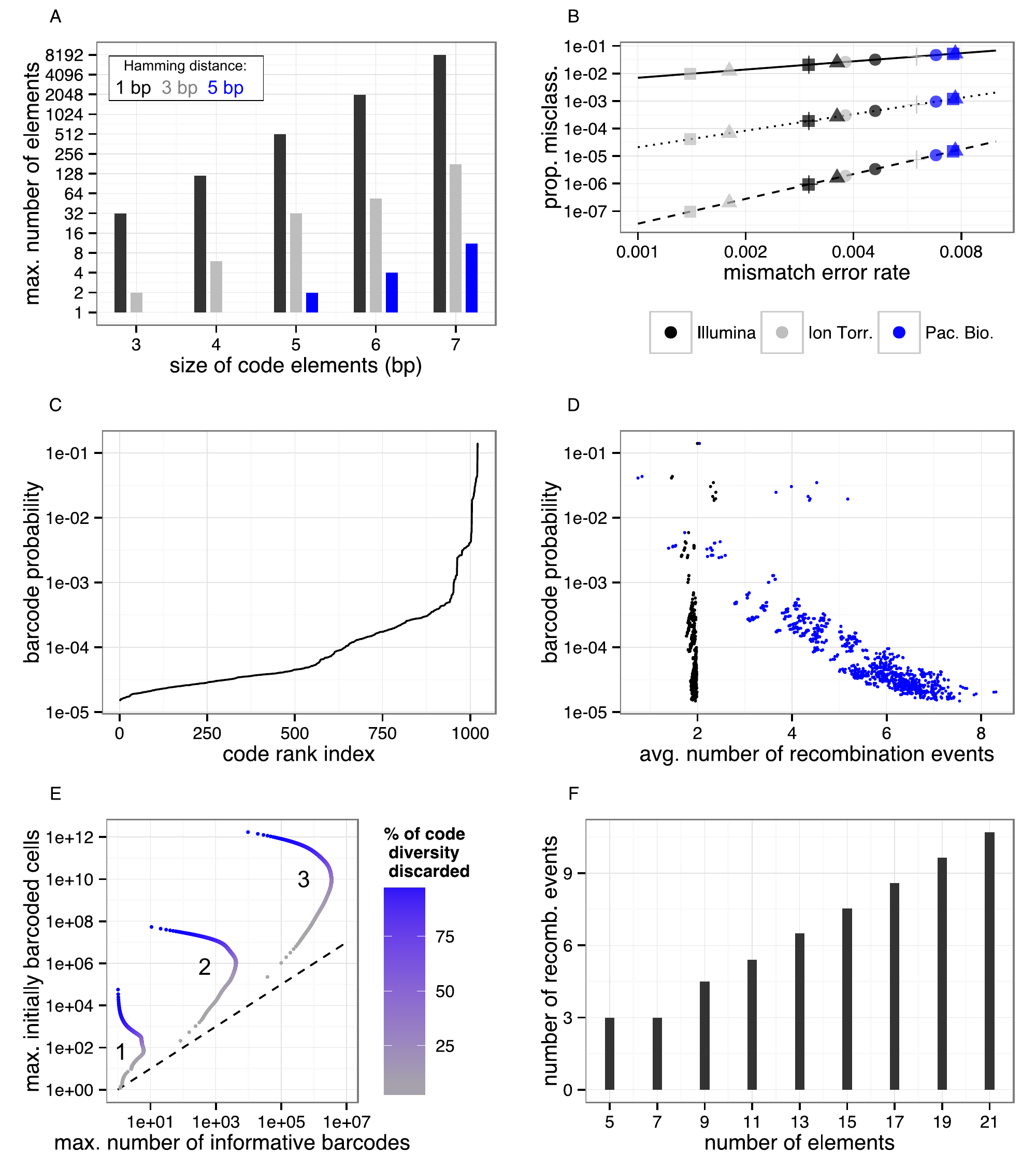}

\caption{Design of code elements and probabilistic features of optimal 
cassettes. \textbf{A}) Computationally determined maximal size of sets
of elements separated by a minimal Hamming distance of 1 bp (black),
3 bp (gray), and 5 bp (blue). To be robust to two sequencing
errors, the minimal distance is 5 bp, which requires $m>4$ bp.
\textbf{B}) Upper bound for the expected
proportion of misclassified elements as a function of empirical DNA
sequencing read error rates \cite{Ross2013Characterizing} for
common sequencing platforms (Illumina, Ion Torrent, Pacific
Biosciences) and different sequence data (P. falciparum ($\bullet$),
E. coli ($\blacktriangle$), R. spha. ($\blacksquare$), H. sapiens
($+$)). The minimal distance that separates the elements is 1 bp
(solid), 3 bp (dotted), and 5 bp (dashed). \textbf{C}) Ranked probabilities
of the 1022 size-stable barcodes from a cassette with 13 elements
generated under constitutive Cre expression. A few codes are relatively
frequent, but the majority are rare. \textbf{D}) Scatter-plot
showing barcode probabilities against the average number of excisions (black)
and the number of inversions (blue) that are generate size-stable barcodes from
a 13 element optimal cassette. \textbf{E})
Number of cells in which a barcode can be induced versus
the number of cells that produce informative codes, for one to three
sequential cassettes, without exceeding 1\% repeated occurrences
in the informative codes. The color represents the percentage of discarded
codes relative to the total code diversity, which can be adjusted
to experimental conditions post acquisition. \textbf{F}) Although
code diversity grows as $O(n^{3})$, the expected number of recombination
events that are needed to generate a size-stable code increases
linearly in $n$.
\label{fig:fig3}}

\end{figure}

Fig. \ref{fig:fig3} C shows the generation probability for
each of the 1022 codes from a cassette with 13 elements.  To produce
this plot, $10^{8}$ barcodes were Monte Carlo generated \textit{in
silico} via sequential recombination of the initial cassette. The
number of times a specific code appeared was recorded, normalized
and sorted. While some codes are relatively frequent, most are rare.
In Fig. \ref{fig:fig3} D, the average number of recombination events
(inversions: blue, excision: black) is plotted as a function of
barcode probability. The number of inversions and barcode probability
are negatively correlated, an indication that rare codes undergo,
on average, more inversions. The number of excisions is close to
two for all codes.

Ideally, each cell is tagged with a unique barcode. As with all
existing barcoding techniques however, 100\% unique barcodes cannot
be guaranteed. What influences the expected number of unique
barcodes is the code diversity $D$, $p_i$, the probability of
code $i$, where $i\in\{1,2,\ldots,D\}$ , and $j$, the total number
of codes that are generated. Using analysis of the generalized
birthday party problem \cite{Koot2012ANALYSIS}, the expected
proportion of unique codes is
\begin{align}
\sum_{i=1}^Dp_{i}(1-p_{i})^{j-1}  
	\approx
	1-(j-1)\sum_{i=1}^Dp_{i}^{2},\label{eq:repeats}
\end{align}
where the numerically convenient approximation on the right hand side
arises from a Taylor expansion around $0$ and is appropriate
if $(j-1)\ll 1/(\max_i p_i)$. Relatively
large $p_{i}$'s negatively affect the expected proportion of unique
codes. For heterogeneous barcode distributions, a natural strategy
is to discard most frequent codes from the analysis. Barcodes that
are included in the final analysis are called informative.

Using the approximation Eq. \eqref{eq:repeats}, in Fig. \ref{fig:fig3}
E we computed the maximum number of cells that can initially be
barcoded versus the number of cells that generate an informative code, for
one to three sequential cassettes (indicated by the numbers 1, 2,
3), with the requirement that no more than 1\% of informative
codes are generated more than once. The color represents the
percentage of discarded codes relative to the total code diversity.
This parameter can be adjusted to meet the needs of a given experiment.
E.g., for three concatenated cassettes with 13 elements
each, $10^{5}$ informative codes that are 99\% unique can be generated
by inducing barcodes in either $10^{6}$ cells and including most
codes or inducing barcodes in $10^{12}$ cells and discarding most
codes from the analysis. These results show that
by discarding frequent codes from the read-out, large numbers of
clones can be confidently tracked, indicating this \emph{in situ}
barcoding is suitable for high-throughput lineage tracing experiments.

\subsubsection*{Number of recombination events to generate barcodes
does not diverge with cassette size }

If Cre is expressed for long enough, Lox cassettes will eventually
become size-stable. The time this will take correlates with the
number of recombination events that separate a stable barcode from
its initial cassette. Below, we estimate this quantity using the
theory of absorbing Markov chains.

In a cassette with $n$ elements, there are $n+1$ Lox sites. The
number of Lox pairs that are flanking $k$ elements is $n+1-k$.  Lox
pairs that have less than three elements between them do not interact
as they are separated by less than the minimal distance. Pairs of
Lox sites that have three or more elements between them are termed
productive. For $n\geq3$ the number of productive pairs is
$\sum_{k=3}^{n}(n+1-k)=(n-1)(n-2)/2$, and the number of productive
pairs, where recombination leads to excision, i.e.
where an even number of elements separates the two sites, is
\begin{align*}
\sum_{3\leq k\leq n: k\text{ even}}(n+1-k)={\displaystyle \frac{(n-1)(n-3)}{4}}
\end{align*}
for $n$ odd.  The probability that a productive pair excises exactly
$k$ elements is given by the ratio of productive
pairs that are separated by $k$ elements to the total number of
productive pairs, i.e.
\begin{align}
P(\textrm{excision of }k\text{ elements})=\frac{n+1-k}{\sum_{k=3}^{n}(n+1-k)}
=  \frac{2(n+1-k)}{(n-1)(n-2)},\label{eq:2}
\end{align}
for $k$ even, $3\leq k\leq n$, otherwise it is zero.  The number
of productive pairs where recombination leads to inversion is (for
$n$ is odd)
\begin{align}
\sum_{3\leq k\leq n,k\textrm{ odd}}(n+1-k)={\displaystyle \frac{(n-1)^{2}}{4}} 
 ,\label{eq:3}
\end{align}
and the probability that interaction of a productive pair leads to
an inversion is
\begin{align}
P(\textrm{inversion})=\displaystyle \frac{2(n-1)^2}{4(n-1)(n-2)}=
{\displaystyle \frac{n-1}{2(n-2)}}  .\label{eq:4}
\end{align}

Equations \eqref{eq:2} - \eqref{eq:4} allow the formulation
of size-stable barcodes as a discrete-time absorbing Markov chain.
The number of elements in the cassette corresponds to its state,
and Eq. \eqref{eq:2} and Eq. \eqref{eq:4} give the transition probabilities
from $n$ to $n-k$, and from $n$ to $n$ elements respectively.  There
are $n-3$ transient and $4$ absorbing states. Absorbing states are
cassettes that have either three, two, one, or zero elements.
Absorbing Markov models are well understood,
and a wealth of theoretical predictions regarding their properties
are available \cite{Grinstead1997Introduction}. 
The fundamental matrix of this Markov Chain is 
\begin{align*}
N=  (I_{n-3}-Q)^{-1}, 
\end{align*} 
where $I_{n-3}$ is an $(n-3)\times(n-3)$ identity matrix, and $Q$
is the transition matrix corresponding to the transient states. The
expected number of recombination events, starting with a cassette
of $n$ elements, until reaching a final code is the $n^\text{th}$
entry of the vector $t=Nc$, where c is a column vector all of whose
entries are 1.

In Fig. \ref{fig:fig3} F, the average number of recombination events
from the initial cassette to final code is shown as a function of
the cassette length. Although code diversity grows as $O(n^{3})$,
the number of recombination events code generation increases linearly
in $n$.

\section*{Discussion}

\subsubsection*{Lox barcode cassettes with code elements of size four}

When we identify the optimal Lox barcode cassette, we focus on code
elements in the regime $5\textrm{ bp}\leq m<24\textrm{ bp}$. These
have maximal size-stable barcodes of three elements that are
insensitive to over-estimation of the minimal Lox interaction
distance.  For $m<5$ bp, size-stable barcodes of four elements are
possible and their maximal code diversity grows as $O(n^{4})$. These
are stable, however, only if the minimal interaction distance between
two Lox sites is greater than $80$ bp, a distance at which interactions
have shown to still be possible \textit{in vivo} in the similar
Flp/FR system \cite{ringrose1999}.

Most interesting is the case $m=4$ bp, which permits correction of
one sequencing error with six code elements that are $3$ bp apart
in both orientations (see gray bars in Fig. \ref{fig:fig3} A). The
upper diversity bound is derived along the same lines as for $m\geq5$
bp (see Fig. \ref{fig:fig4} E for possible stable codes), which gives
\begin{align*}
  48\dbinom{n_{r}}{1}\dbinom{n_{b}}{3}
	+64\dbinom{n_{r}}{2}\dbinom{n_{g}}{1}\dbinom{n_{b}}{1}
	+12\dbinom{n_{r}}{1}\dbinom{n_{b}}{2}
	+16\dbinom{n_{r}}{2}\dbinom{n_{g}}{1}
	+4\dbinom{n_{r}}{1}\dbinom{n_{b}}{1}
	+2\dbinom{n_{r}}{1}.
\end{align*}
To maximize usage of the 6 code elements, we start with a cassette
that has six red, five green and six blue blocks, i.e.
$\{n_{r},n_{g},n_{b}\}=\{6,5,6\}$. This gives an upper diversity
bound of 36996 barcodes. As confirmed by simulations, this upper
bound is attained by a cassette with inverted flanking sites in
which the first 11 Lox sites are alternating, and the remaining
sites, except the last, are oriented in the same direction as the
first Lox site (Fig. \ref{fig:fig4} F). Under constitutive Cre
expression, barcodes with four elements can still undergo inversions,
and the effective code diversity is 19,716.

Careful measurements will be needed to determine whether Lox sites
at a distance of 80 bp still interact. If they don't, the cassette
shown in Fig. \ref{fig:fig4} F with $m=4$ bp represents an interesting
alternative to the design described in the main text, as with less
elements it reaches higher code diversity, but at the cost of less
robustness to sequencing error and hence barcode readout fidelity.

\subsubsection*{Higher order Lox interactions}

Single recombination events always involve exactly two Lox sites.
However nothing except DNA flexibility prevents several pairs of
Lox sites to interact simultaneously. The rate at which pairs of
Lox sites bind depends on the number of Lox sites and the kinetic
rates of Lox-Lox complexes. \textit{In vitro}, the latter appear
stable \cite{Pinkney2012Capturing} and with the potentially large
number of Lox sites in the barcode cassettes, make simultaneous
interactions a plausible possibility.

Higher order Lox interactions lead to previously unreported, and
in certain cases novel, recombination products (Fig. \ref{fig:fig4}
C). For example, simultaneous interactions of two overlapping pairs
of Lox sites oriented in the same direction do not result in excision,
but in a reordering of the sequences between the sites. Similarly,
if pairs are inverted, simultaneous recombinations do not invert
but excise the sequence between the outermost sites.

For the alternating cassette and $n\geq7$, multiple concurrent Lox
interactions do not generate additional codes as the upper code
diversity bound is already attained. Therefore our results on Lox
barcode design and code elements remain unchanged in the presence
of higher order Lox interactions. What changes is the distribution
over barcodes, which flattens in the tail if more than one Lox pair
recombines at a time (Fig. \ref{fig:fig4} D).

\subsubsection*{Transient Cre expression. }

Code diversity strongly depends on the number of elements in
size-stable barcodes. If Cre is expressed constitutively, size-stable
barcodes with code elements of size $m\geq5$ bp have a maximum of
three elements. Another possibility is to create transient Cre
activity rather than constitutive.

A well tested system that provides temporal control over Cre activity
is tamoxifen inducible CreEr \cite{Nagy2000Cre}. In the presence
of tamoxifen, the fusion protein CreEr, which is normally located
in the cytoplasm, is transported into the nucleus, where it can bind
to Lox sites and induce recombination. Depending on the duration
of Cre activation and its efficiency, stable sequences with more than
three elements are likely to be generated from a Lox barcode
cassette.  Although most of these sequences are stable only in the
absence of Cre, in this section we make no distinction between these
and the size-stable barcodes defined earlier.

\begin{figure}
\centering\includegraphics[scale=0.65]{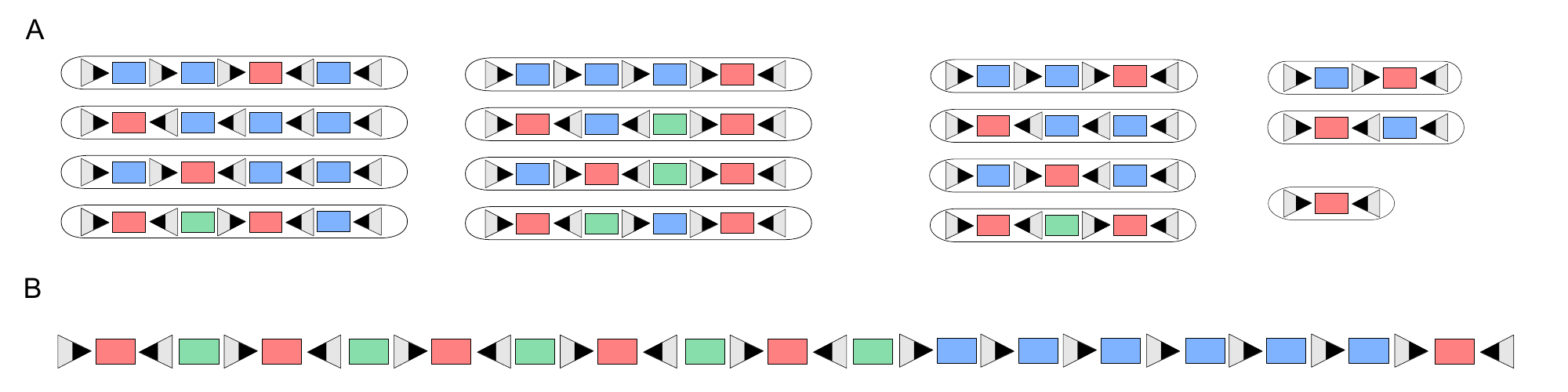}\\
\includegraphics[scale=0.65]{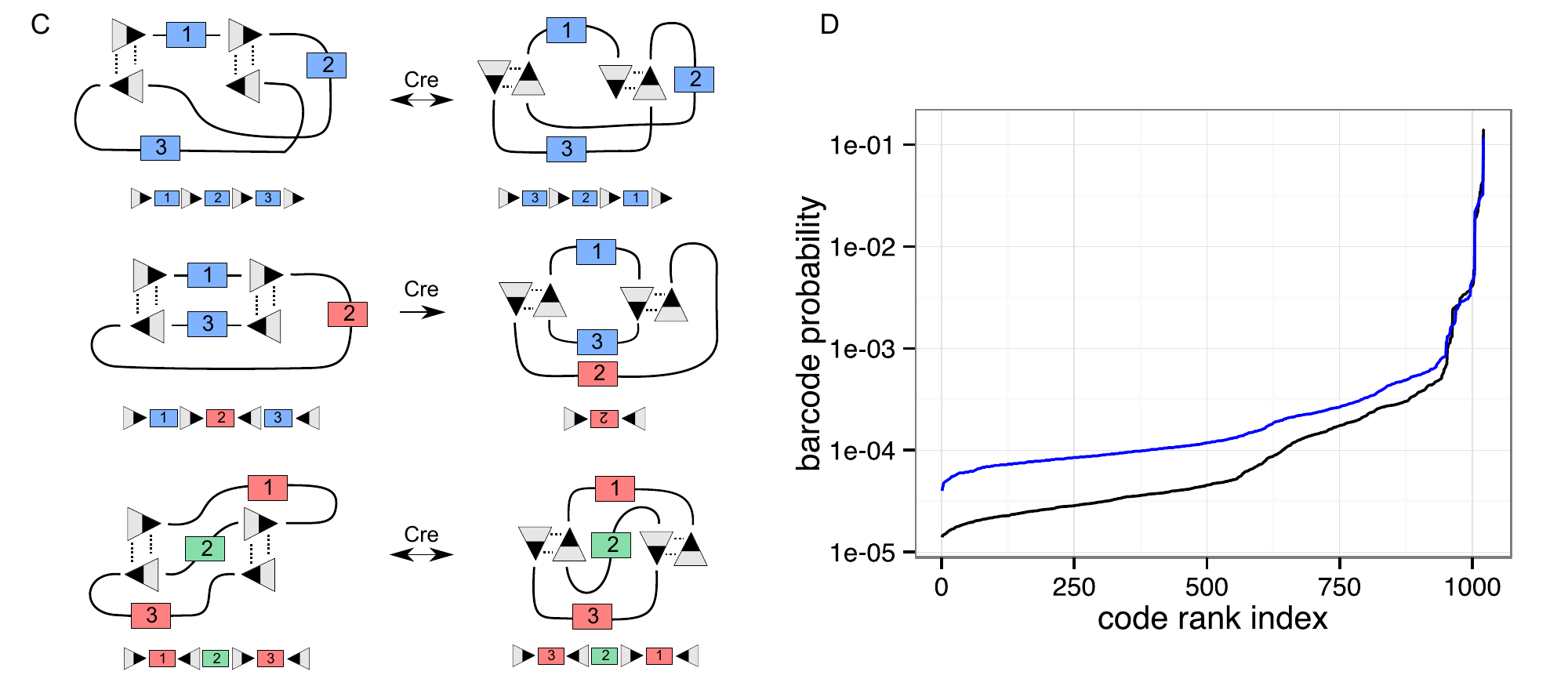}\\
\includegraphics[scale=0.65]{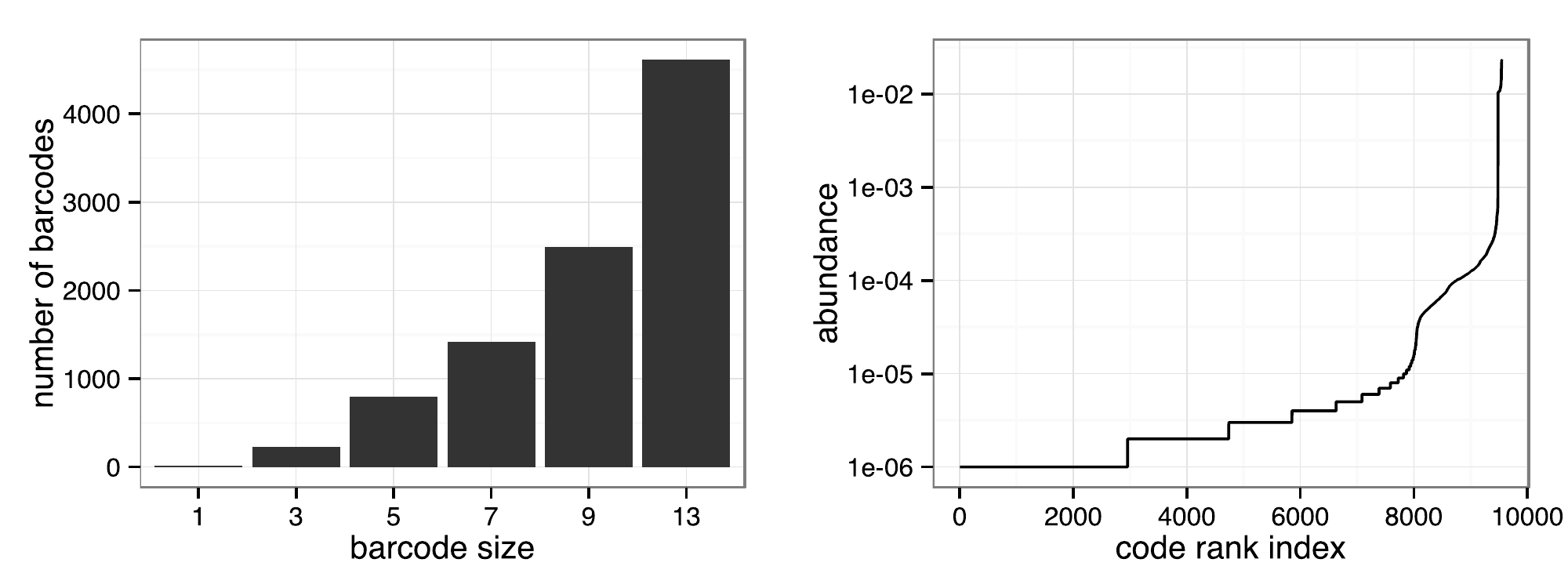}

\caption{Short code elements, higher order Lox interactions, and transient
Cre activation. \textbf{A}) Possible size-stable barcodes if $m<5$
bp. \textbf{B}) Cassette with 17 elements and $m=4$ bp that attains
an effective code diversity of 19,716 barcodes if the minimal Lox
interaction distance is greater than $80$ bp. \textbf{C}) If two
ore more pairs of lox sites recombine simultaneously, unexpected
recombination products can occur. \textbf{D}) Estimated barcode
distribution if two Lox pairs can interact simultaneously (blue).
The distribution becomes flatter at the lower end, implying that
rare codes are more likely than if recombination events only occur
sequentially (black). \textbf{ E}) Mimicking a short Cre activation
pulse in a population of a million cells carrying a 13 element Lox
cassette, the number of recombination events is assumed Poisson
distributed with mean $1$. After the pulse many barcodes have not
experienced any excisions. \textbf{F}) Code abundance after the
pulse. Almost $10^{4}$ distinct barcodes are generated, with 
~30\% being generated once.  \label{fig:fig4}}

\end{figure}

Fig. \ref{fig:fig4} A shows barcode probabilities after activation of
CreEr in $10^{6}$ cells with an optimal Lox cassette of size 13.
The number of recombination events induced by transient CreEr activity
is assumed Poisson with mean one. About $10^{4}$ distinct barcodes
are generated, and 30\% of these appear only once. Although promising
in terms of code diversity, it should be noted that potential
drawbacks of this approach are the length of the barcodes (leading
to more involved code sequencing), leakiness of CreEr into the
nucleus in non-induced cells \cite{Kretzschmar2012Lineage}, and the
relatively long half-life of tamoxifen \cite{Reinert2012TamoxifenInduced}.

Existing cellular barcoding approaches have already lead to significant
biological discoveries and so new approaches that overcome their
shortcomings are inherently desirable. Here we have established
that using Cre Lox, it would be feasible to create an \emph{in
situ}, triggerable barcoding system with sufficient diversity to
label a whole mouse, and propose this as a system for experimental
implementation.

{\bf Acknowledgments:}
Ton Schumacher (Netherlands Cancer Institute) for informative
discussions.

{\bf Funding:} 
T.W., S.N and K.D. were supported by Human Frontier Science Program
grant RGP0060/2012. K.D. was also supported by Science Foundation
Ireland grant 12 IP 1263.



\begin{thebibliography}{10}

\bibitem{buchholz2013}
Buchholz VR, Flossdorf M, Hensel I, Kretschmer L, Weissbrich B, Gr{\"a}f P,
  et~al.
\newblock Disparate individual fates compose robust {CD8+ T cell} immunity.
\newblock Science. 2013;340(6132):630--635.

\bibitem{Gerlach2010One}
Gerlach C, van Heijst JWJ, Swart E, Sie D, Armstrong N, Kerkhoven RM, et~al.
\newblock {One naive T cell, multiple fates in CD8+ T cell differentiation}.
\newblock J Exp Med. 2010;207(6):1235--1246.

\bibitem{Verovskaya2013Heterogeneity}
Verovskaya E, Broekhuis MJ, Zwart E, Ritsema M, van Os R, de~Haan G, et~al.
\newblock {Heterogeneity of young and aged murine hematopoietic stem cells
  revealed by quantitative clonal analysis using cellular barcoding.}
\newblock Blood. 2013;122(4):523--532.

\bibitem{Naik2013Diverse}
Naik SH, Peri\'{e} L, Swart E, Gerlach C, van Rooij N, de~Boer RJ, et~al.
\newblock {Diverse and heritable lineage imprinting of early haematopoietic
  progenitors}.
\newblock Nature. 2013;496(7444):229--232.

\bibitem{perie15}
Peri{\'e} L, Duffy KR, Kok L, de~Boer RJ, Schmacher TN.
\newblock The branching point in erythro-myeloid differentiation.
\newblock Cell. 2015;163(7):1655--1662.

\bibitem{Johnson2015Singlecell}
Johnson MB, Wang PP, Atabay KD, Murphy EA, Doan RN, Hecht JL, et~al.
\newblock {Single-cell analysis reveals transcriptional heterogeneity of neural
  progenitors in human cortex}.
\newblock Nat Neurosci. 2015;18(5):637--646.

\bibitem{Yagi2013Genetic}
Yagi T.
\newblock {Genetic basis of neuronal individuality in the mammalian brain.}
\newblock J Neurogenet. 2013;27(3):97--105.

\bibitem{Zeisel2015Brain}
Zeisel A, Mu\~{n}oz Manchado AB, Codeluppi S, L\"{o}nnerberg P, La~Manno G,
  Jur\'{e}us A, et~al.
\newblock Cell types in the mouse cortex and hippocampus revealed by
  single-cell {RNA}-seq.
\newblock Science. 2015;347(6226):1138--1142.

\bibitem{NolanStevaux2013Measurement}
Nolan-Stevaux O, Tedesco D, Ragan S, Makhanov M, Chenchik A, Ruefli-Brasse A,
  et~al.
\newblock Measurement of cancer cell growth heterogeneity through lentiviral
  barcoding identifies clonal dominance as a characteristic of in vivo Ttmor
  engraftment.
\newblock PLoS One. 2013;8(6):e67316+.

\bibitem{Bhang2015Studying}
Bhang HeC, Ruddy DA, Krishnamurthy~Radhakrishna V, Caushi JX, Zhao R, Hims MM,
  et~al.
\newblock {Studying clonal dynamics in response to cancer therapy using
  high-complexity barcoding}.
\newblock Nat Med. 2015;21(5):440--448.

\bibitem{Klauke2015Tracing}
Klauke K, Broekhuis MJC, Weersing E, Dethmers-Ausema A, Ritsema M, Gonz\'{a}lez
  MV, et~al.
\newblock Tracing dynamics and clonal heterogeneity of Cbx7-induced leukemic
  stem cells by cellular barcoding.
\newblock Stem Cell Reports. 2015;4(1):74--89.

\bibitem{rohr2014single}
Rohr JC, Gerlach C, Kok L, Schumacher TN.
\newblock Single cell behavior in T cell differentiation.
\newblock Trends Immunol. 2014;35(4):170--177.

\bibitem{Duffy2012Intracellular}
Duffy KR, Hodgkin PD.
\newblock {Intracellular competition for fates in the immune system}.
\newblock Trends Cell Biol. 2012;22(9):457--464.

\bibitem{Hawkins09}
Hawkins ED, Markham JF, McGuinness LP, Hodgkin PD.
\newblock A single-cell pedigree analysis of alternative stochastic lymphocyte
  fates.
\newblock Proc Natl Acad Sci USA. 2009;106(32):13457--13462.

\bibitem{Rieger2009}
Rieger MA, Hoppe PS, Smejkal BM, Eitelhuber AC, Schroeder T.
\newblock Hematopoietic cytokines can instruct lineage choice.
\newblock Science. 2009;325(5937):217--218.

\bibitem{Gomes11}
Gomes FL, Zhang G, Carbonell F, Correa JA, Harris WA, Simons BD, et~al.
\newblock Reconstruction of rat retinal progenitor cell lineages in vitro
  reveals a surprising degree of stochasticity in cell fate decisions.
\newblock Development. 2011;138(2):227--235.

\bibitem{Giurumescu2012Quantitative}
Giurumescu CA, Kang S, Planchon TA, Betzig E, Bloomekatz J, Yelon D, et~al.
\newblock {Quantitative semi-automated analysis of morphogenesis with
  single-cell resolution in complex embryos}.
\newblock Development. 2012;139(22):4271--4279.

\bibitem{Duffy12}
Duffy KR, Wellard CJ, Markham JF, Zhou JHS, Holmberg R, Hawkins ED, et~al.
\newblock Activation-induced {B} cell fates are selected by intracellular
  stochastic competition.
\newblock Science. 2012;335(6066):338--341.

\bibitem{richards2013}
Richards JL, Zacharias AL, Walton T, Burdick JT, Murray JI.
\newblock A quantitative model of normal Caenorhabditis elegans embryogenesis
  and its disruption after stress.
\newblock Dev Biol. 2013;374(1):12--23.

\bibitem{etzrodt2014quantitative}
Etzrodt M, Endele M, Schroeder T.
\newblock Quantitative single-cell approaches to stem cell research.
\newblock Cell stem cell. 2014;15(5):546--558.

\bibitem{cohen2014extracting}
Cohen AR.
\newblock Extracting meaning from biological imaging data.
\newblock Mol Biol Cell. 2014;25(22):3470--3473.

\bibitem{Lu2011Tracking}
Lu R, Neff NF, Quake SR, Weissman IL.
\newblock {Tracking single hematopoietic stem cells in vivo using
  high-throughput sequencing in conjunction with viral genetic barcoding}.
\newblock Nat Biotech. 2011;29(10):928--933.

\bibitem{Naik2014Cellular}
Naik SH, Schumacher TN, Peri\'{e} L.
\newblock {Cellular barcoding: a technical appraisal}.
\newblock Exp Hematol. 2014;42(8):598--608.

\bibitem{perie14}
Peri{\'e} L, Hodgkin PD, Naik SH, Schumacher TN, de~Boer RJ, Duffy KR.
\newblock Determining lineage pathways from cellular barcoding experiments.
\newblock Cell Rep. 2014;6(4):617--624.

\bibitem{Sun2014Clonal}
Sun J, Ramos A, Chapman B, Johnnidis JB, Le L, Ho YJ, et~al.
\newblock {Clonal dynamics of native haematopoiesis}.
\newblock Nature. 2014;514(7522):322--327.

\bibitem{Zador2012Sequencing}
Zador AM, Dubnau J, Oyibo HK, Zhan H, Cao G, Peikon ID.
\newblock {Sequencing the Connectome}.
\newblock PLoS Biol. 2012;10(10):e1001411+.

\bibitem{Wei2012Exactly}
Wei Y, Koulakov AA.
\newblock An exactly solvable model of random site-specific recombinations.
\newblock Bull Math Biol. 2012;74(12):2897--2916.

\bibitem{Peikon2014In}
Peikon ID, Gizatullina DI, Zador AM.
\newblock {In vivo generation of DNA sequence diversity for cellular
  barcoding}.
\newblock Nucleic Acids Res. 2014;42(16):e127.

\bibitem{Livet2007Transgenic}
Livet J, Weissman TA, Kang H, Draft RW, Lu J, Bennis RA, et~al.
\newblock Transgenic strategies for combinatorial expression of fluorescent
  proteins in the nervous system.
\newblock Nature. 2007;450(7166):56--62.

\bibitem{Hoess1985Formation}
Hoess R, Wierzbicki A, Abremski K.
\newblock {Formation of small circular DNA molecules via an in vitro
  site-specific recombination system.}
\newblock Gene. 1985;40(2-3):325--329.

\bibitem{ringrose1999}
Ringrose L, Chabanis S, Angrand PO, Woodroofe C, Stewart AF.
\newblock Quantitative comparison of {DNA} looping in vitro and in vivo:
  chromatin increases effective {DNA} flexibility at short distances.
\newblock The EMBO Journal. 1999;18(23):6630--6641.

\bibitem{Pinkney2012Capturing}
Pinkney JN, Zawadzki P, Mazuryk J, Arciszewska LK, Sherratt DJ, Kapanidis AN.
\newblock {Capturing reaction paths and intermediates in Cre-loxP recombination
  using single-molecule fluorescence.}
\newblock Proc Natl Acad Sci USA. 2012;109(51):20871--20876.

\bibitem{Parrish2011BAC}
Parrish M, Unruh J, Krumlauf R.
\newblock {BAC} modification through serial or simultaneous use of {CRE/Lox}
  technology.
\newblock J Biomed Biotechnol. 2011;2011:1--12.

\bibitem{Nagy2000Cre}
Nagy A.
\newblock {Cre recombinase: the universal reagent for genome tailoring.}
\newblock Genesis. 2000;26:99--109.

\bibitem{Blattman2002Estimating}
Blattman JN, Antia R, Sourdive DJD, Wang X, Kaech SM, Murali-Krishna K, et~al.
\newblock Estimating the precursor frequency of naive antigen-specific {CD8 T}
  cells.
\newblock J Exp Med. 2002;195(5):657--664.

\bibitem{Colvin2004Murine}
Colvin GA, Lambert JF, Abedi M, Hsieh CC, Carlson JE, Stewart FM, et~al.
\newblock {Murine marrow cellularity and the concept of stem cell competition:
  geographic and quantitative determinants in stem cell biology}.
\newblock Leukemia. 2004;18(3):575--583.

\bibitem{Sternberg1981Bacteriophage}
Sternberg N, Hamilton D, Hoess R.
\newblock {Bacteriophage P1 site-specific recombination. II. Recombination
  between loxP and the bacterial chromosome.}
\newblock J Mol Biol. 1981;150(4):487--507.

\bibitem{Hamilton1984Sitespecific}
Hamilton DL, Abremski K.
\newblock {Site-specific recombination by the bacteriophage P1 lox-Cre system.
  Cre-mediated synapsis of two lox sites.}
\newblock J Mol Biol. 1984;178(2):481--486.

\bibitem{Guo1997Structure}
Guo F, Gopaul DN, Van~Duyne GD.
\newblock {Structure of Cre recombinase complexed with DNA in a site-specific
  recombination synapse}.
\newblock Nature. 1997;389(6646):40--46.

\bibitem{Bystrykh2012Counting}
Bystrykh LV, Verovskaya E, Zwart E, Broekhuis M, de~Haan G.
\newblock {Counting stem cells: methodological constraints}.
\newblock Nat Meth. 2012;9(6):567--574.

\bibitem{Cover1991Elements}
Cover TM, Thomas JA.
\newblock {Elements of Information Theory}.
\newblock Wiley-Interscience; 1991.

\bibitem{Li2008Mapping}
Li H, Ruan J, Durbin R.
\newblock {Mapping short DNA sequencing reads and calling variants using
  mapping quality scores.}
\newblock Genome Res. 2008;18(11):1851--1858.

\bibitem{Ross2013Characterizing}
Ross MG, Russ C, Costello M, Hollinger A, Lennon NJ, Hegarty R, et~al.
\newblock {Characterizing and measuring bias in sequence data}.
\newblock Genome Biol. 2013;14(5):R51+.

\bibitem{Koot2012ANALYSIS}
Koot MR, Mandjes M.
\newblock The analysis of singletons in generalized birthday problems.
\newblock Probab Eng Inform Sc. 2012;26(2):245--262.

\bibitem{Grinstead1997Introduction}
Grinstead CM, Snell JL.
\newblock {Introduction to Probability}.
\newblock 2nd ed. American Mathematical Society; 1997.

\bibitem{Kretzschmar2012Lineage}
Kretzschmar K, Watt FM.
\newblock {Lineage Tracing}.
\newblock Cell. 2012;148(1-2):33--45.

\bibitem{Reinert2012TamoxifenInduced}
Reinert RB, Kantz J, Misfeldt AA, Poffenberger G, Gannon M, Brissova M, et~al.
\newblock Tamoxifen-induced {Cre-loxP} recombination is prolonged in pancreatic
  islets of adult mice.
\newblock PLoS One. 2012;7(3):e33529+.

\end{thebibliography}
\end{document}